# In situ growth regime characterization of cubic GaN using reflection high energy electron diffraction

J. Schörmann,[a] S. Potthast, D. J. As, and K. Lischka
*Department of Physics, University of Paderborn, Warburger Strasse 100, D-33095 Paderborn, Germany*

Cubic GaN layers were grown by plasma-assisted molecular beam epitaxy on 3$C$-SiC (001) substrates. *In situ* reflection high energy electron diffraction was used to quantitatively determine the Ga coverage of the GaN surface during growth. Using the intensity of the electron beam as a probe, optimum growth conditions of $c$-GaN were found when a 1 ML Ga coverage is formed at the surface. 1 $\mu$m thick $c$-GaN layers had a minimum surface roughness of 2.5 nm when a Ga coverage of 1 ML was established during growth. These samples revealed also a minimum full width at half maximum of the (002) rocking curve. © *2007 American Institute of Physics*.
[DOI: 10.1063/1.2432293]

Group III-nitrides crystallize in the stable wurtzite structure or in the metastable zinc blende structure. An important difference between these material modifications is the presence of strong internal electric fields in hexagonal (wurtzite) III-nitrides grown along the polar (0001) $c$ axis, while these "built-in" fields are absent in cubic (zinc blende) III-nitrides. The zinc blende GaN polytype is metastable and can only be grown in a very narrow window of process conditions.[1] However, the use of nearly lattice matched, freestanding high quality 3$C$-SiC (001) substrates let to substantial improvements of the crystal quality of $c$-III-nitrides.[2]

As an important step to improve the GaN surface morphology in a systematic way, it is essential to understand the surface structure and the underlying growth process on an atomic scale. In particular, the kinetic processes of adsorption and desorption on the surface are considered as key parameters that govern the surface morphology, incorporation kinetics, and consecutively the overall material quality. In molecular beam epitaxy (MBE) of hexagonal GaN, two-dimensional surfaces are commonly achieved under Ga-rich conditions, with theoretical[3] and experimental[4,5] evidences suggesting that the growth front is stabilized by a metallic Ga adlayer.

In this letter, we present a study of the Ga adsorption on cubic (001) GaN surfaces by reflection high energy electron diffraction (RHEED). We developed a method which allows measuring the Ga coverage of the GaN surface during growth with submonolayer accuracy. We show quantitatively that a 1 ML coverage favors two-dimensional growth and yields $c$-GaN layers with a minimum surface roughness.

All $c$-GaN layers were grown on freestanding 3$C$-SiC (001) substrates by plasma assisted molecular beam epitaxy. An *Oxford Applied Research* HD25 radio frequency plasma source was used to provide activated nitrogen atoms. Gallium was evaporated from Knudsen cells. Cubic GaN layers were deposited at 720 °C directly on 3$C$-SiC substrates. The adsorption and desorption of metal (Ga) layers on the $c$-GaN surface were investigated using the intensity of a reflected high energy electron beam (short "RHEED intensity") as a probe. After closing all shutters the GaN surface was exposed to different Ga fluxes for a certain time. The substrate temperature was kept at $T_S = 720$ °C. The transient of the RHEED intensity of the (0,0) streak was recorded. This procedure was also used to measure the Ga coverage during $c$-GaN growth. Structural characterization was carried out by high resolution x-ray diffraction and atomic force microscopy (AFM).

The optimum conditions for the epitaxial growth of $c$-GaN are mainly determined by two parameters, the surface stoichiometry and the substrate temperature.[1] Both parameters are interrelated; therefore an *in situ* control of substrate temperature and surface stoichiometry is highly desirable. The study of the surface reconstruction by RHEED was one of the key issues in understanding the $c$-III-nitride growth.[1,6,7] First principles calculations by Neugebauer *et al.*[8] show that all energetically favored surface modifications of the nonpolar (001) $c$-GaN surface are Ga stabilized and therefore optimum growth conditions are expected under slightly Ga-rich conditions.

We found that for a submonolayer Ga coverage of a $c$-GaN crystal surface, the RHEED intensity decreases. By calibrating the reflectivity drop the Ga coverage can be measured. The calibration has been performed as follows: The $c$-GaN layers, which were used for the RHEED experiments, were grown under Ga-rich conditions with a coverage of about 1 ML as described below. All experiments used for the RHEED intensity observation were started on a Ga-free $c$-GaN surface established by a 5 min growth interruption. The adsorption and desorption of gallium on a $c$-GaN (001) surface were observed by measuring the RHEED intensity. Figure 1 shows a plot of the RHEED intensity versus time which has been measured after exposing $c$-GaN at 720 °C to Ga fluxes of $4.3 \times 10^{14}$ and $7.5 \times 10^{14}$ cm$^{-2}$ s$^{-1}$, respectively. After opening the Ga shutter we observe a steep decrease of the RHEED intensity between $I_0$ and the kink position ($I_k$) indicated in Fig. 1. The gradient of the intensity drop is related to the impinging Ga flux. The further decrease of the RHEED intensity has a different gradient. After closing the Ga shutter an increase of the RHEED intensity is observed which is due to desorption of accumulated Ga. Again two different slopes, which change at $I_k$, are observed. Finally, the RHEED intensity reaches the starting value $I_0$, indicating a Ga-free surface.

---

[a]Electronic mail: li_jsch@physik.upb.de

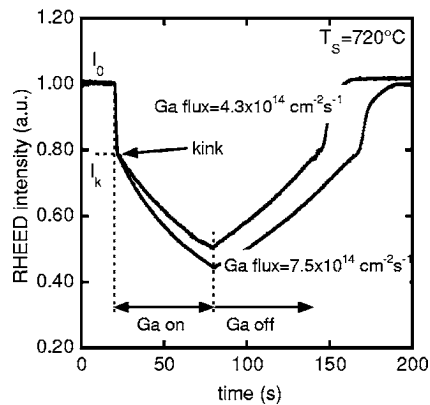

FIG. 1. Intensity of a reflected high energy electron beam (RHEED intensity) vs time measured during the evaporation of Ga onto *c*-GaN at a substrate temperature of 720 °C. The Ga fluxes are indicated.

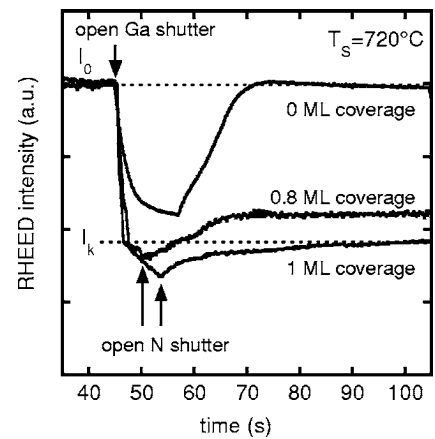

FIG. 3. RHEED intensity transient measured during the growth of *c*-GaN, which started after opening the *N* source. The RHEED intensity measured during growth yields the amount of excess Ga (indicated in the figure) on the *c*-GaN surface. The Ga fluxes are $4.4 \times 10^{14}$, $3.2 \times 10^{14}$, and $1.2 \times 10^{14}$ cm$^{-2}$ for the coverages of 1, 0.8, and 0 ML, respectively.

Figure 2 shows the RHEED intensity just after opening of the Ga shutter on an extended time scale. A linear decrease of the RHEED intensity up to the kink position is observed. Using the known value of the Ga flux and the time $\Delta t_k$ it takes for the RHEED intensity to drop to $I_k$, we are able to calculate the amount of adsorbed gallium. Assuming that re-evaporation of Ga can be neglected, we get a total number of adsorbed Ga atoms from the flux times $\Delta t_k$ product. In both cases this product is about $9.8 \times 10^{14}$ cm$^{-2}$, which is equal the number of atoms on a GaN surface (lattice constant of 4.52 Å), indicating that 1 ML of Ga is adsorbed during $\Delta t_k$. This result was confirmed using other Ga fluxes exceeding $3 \times 10^{14}$ cm$^{-2}$ s$^{-1}$. Since $I_0$ is the reflectivity of the GaN surface and $I_k$ is the reflectivity of GaN covered by 1 ML Ga, and the drop of the RHEED intensity in the time interval $\Delta t_k$ is linear, the Ga coverage between 0 and 1 ML can be inferred from the measured intensity drop by linear interpolation. The decrease of the RHEED intensity below $I_k$ is most likely due to further accumulation of Ga and thereby a modification of the Ga adlayer surface; however, it is not proportional to the amount of adsorbed Ga. For this reason our method can only be used to measure the Ga coverage between 0 and 1 ML, respectively.

For Ga fluxes less than $3 \times 10^{14}$ cm$^{-2}$ s$^{-1}$ the situation is different. For these fluxes it is not possible to define a kink position. The RHEED intensity drops to a certain value and saturates. We suppose that then desorption of Ga cannot be further neglected. However, we found that these flux values are not relevant for *c*-GaN epitaxy.

The Ga coverage during *c*-GaN growth was measured as follows. Figure 3 shows the RHEED intensity measured after opening the Ga shutter. After the RHEED intensity reached the kink position the *N* shutter was opened too. Then we observe an increase of the intensity which is due to the formation of *c*-GaN. During further growth the RHEED intensity saturates. From the saturation value the Ga coverage can be calculated using $I_k$ as a reference. This procedure allows measuring the Ga coverage in the range between 0 and 1 ML with an accuracy of 0.1 ML.

Figure 4 shows the root-mean-square (rms) roughness measured by a $5 \times 5$ $\mu$m$^2$ AFM scan of several *c*-GaN layers versus the Ga flux used during MBE. The nitrogen flux was almost identical for all samples. The corresponding values of the Ga coverage during growth, as measured by the procedure described above, are also included in Fig. 4. Only values below 1 ML can be measured. Minimum roughness is obtained with a 1 ML Ga coverage during growth. This is in contrast to what has been observed with *h*-GaN, where the optimum growth conditions with regard to surface morphology are related to the formation of a Ga bilayer (*c* plane,[9,10]) or a trilayer (*m* plane,[5]), respectively.

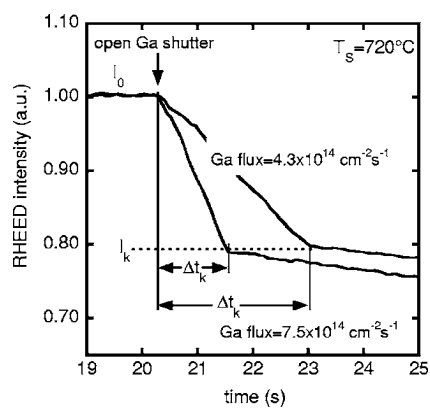

FIG. 2. RHEED intensity transients for two different Ga fluxes as indicated in the figure. After a transition time $\Delta t_k$ a kink in the transients is observed. Calculating the amount of Ga absorbed (flux times $\Delta t_k$) we find that the kink indicates the adsorption of 1 ML Ga on the *c*-GaN surface.

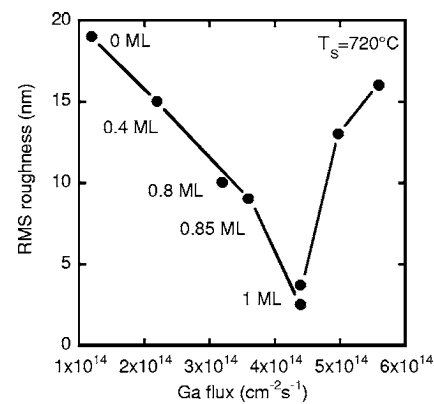

FIG. 4. rms roughness of *c*-GaN layers measured by $5 \times 5$ $\mu$m$^2$ AFM scans vs Ga flux during growth. The corresponding values of the Ga coverage during growth are also included. Minimum roughness is obtained with an excess coverage of 1 ML. The line is a guide for the eyes.

It has variously been suggested that excess Ga acts as surfactant during the epitaxy of hexagonal GaN.[9,11,12] We believe that our data shown in Fig. 4 clearly demonstrate that this effect exists also on the (001) surface of *c*-GaN. The width of the (002) x-ray rocking curve measured in the double axis configuration of 1 μm thick *c*-GaN layers grown with 1 ML coverage is about 16 arc min. Among our *c*-GaN layers with equal thickness, 16 arc min is a minimum value. Gallium fluxes which are equivalent to a Ga coverage exceeding 1 ML lead to a pronounced increase of the roughness and the full width at half maximum of the x-ray rocking curve.

In summary we have studied the adsorption of Ga on *c*-GaN using *in situ* reflection high energy electron diffraction. The drop of the RHEED intensity after the growth was started has been calibrated to allow direct measurements of the Ga coverage at growth conditions with an accuracy of 0.1 ML. The roughness of cubic GaN grown by MBE on free-standing 3*C*-SiC (001) substrates was significantly reduced by growth under controlled Ga-excess conditions. A minimum rms roughness of 2.5 nm was achieved using a Ga coverage of 1 ML during *c*-GaN growth. Cubic GaN layers grown under these conditions on 3*C*-SiC substrates have narrow x-ray (002) rocking curves (16 arc min) indicating also a low density of extended defect in these layers. Our data show that the epitaxy of *c*-GaN with high structural quality is only possible when the amount of excess Ga on the surface is monitored with high accuracy during growth.

The authors would like to thank H. Nagasawa and M. Abe from SiC Development Center, HOYA Corporation, for supplying the 3*C*-SiC substrates.